\documentstyle[aps, preprint]{revtex}
\begin{document}
\tightenlines
\title{Inflationary cosmology from STM theory of gravity}
\author{Mauricio Bellini\footnote{E-mail address: mbellini@mdp.edu.ar}}
\address{Consejo Nacional de Investigaciones Cient\'{\i}ficas y
T\'ecnicas (CONICET)\\
and\\
Departamento de F\'{\i}sica, Facultad de Ciencias
Exactas y Naturales,
Universidad Nacional de Mar del Plata,
Funes 3350, (7600) Mar del Plata, Buenos Aires, Argentina.}
\maketitle
\begin{abstract}
I study the power-law
and de Sitter expansions for the universe during inflation from
the STM theory of gravity.
In a de Sitter expansion the additional
dimension is related to the cosmological constant $\Lambda = 3/\psi^2$.
I find from experimental data that the mass of the inflaton field
is $m^2=2/(3\psi^2)$. The interesting in this case is that the inflaton
field fluctuations are related to the fifth coordinate.
In power-law expansion, the fifth coordinate ($\psi$)
appears to be a dimensionless constant. Here, 
the $\psi$-value depends on the initial conditions.
I find the 5D line element for this inflationary expansion, which is
a function of the classical component of the
inflaton. But the more important result here obtained is that in 
both cases there isn't dynamical compactification during
inflation.
\end{abstract}
\vskip .2cm                             
\noindent
Pacs numbers: 98.80.Cq, 04.20.Jb, 11.10.kk\\
\vskip 1cm
\section{Introduction}

The extra dimension is already known to be of potential importance for
cosmology. There is a class of five-dimensional (5D) cosmological models
which reduce to the usual four-dimensional ones, on hypersurfaces defined
by setting the value of the extra coordinate equal to a constant\cite{1}.
In these models, physical properties such as the mean energy density
and pressure of matter are well defined consequences of how the extra
coordinate enters the metric\cite{2}. That is, matter is explained
as the consequence of geometry in five dimensions. The physics of this
follows from a mathematical result, which is that Einstein's equations
of general relativity with matter are a subset of the Kaluza-Klein (KK)
equations in apparent vacuum.
Hence, cosmology is a subject of pure geometry in
five-dimensional Kaluza-Klein gravity known as
space-time-matter (STM) theory. This theory is supported by more
local astrophysics\cite{2'}.
There is a canonical class of five-dimensional metrics,
most often discussed in the form due to Gross with Perry\cite{3} and Davidson
with Owen\cite{4}.
During the last years there were many attempts to construct a consistent
brane cosmology\cite{.}. However, after a detailed investigation of this
possibility a series of no-go theorems have been proven\cite{..}
There is a
main difference between STM and Brane-World (BW)\cite{...}
formalisms. In the STM theory of gravity we dont need to
insert any matter into the 5D manifold by hand, as is commonly
done in the BW formalism. In the STM
theory of gravity the 5D metric is an exact solution of the 5D field
equations in apparent vacuum\cite{5}. The interesting here, is that
matter appears in four dimensions without any dimensional compactification,
but induced by the 5D vacuum conditions.

On the other hand, inflationary cosmology is one of the most
reliable concepts in modern cosmology.
The first model of inflation was proposed by A. Starobinsky in 1979\cite{1'}.
A much simpler inflationary model with a clear motivation was developed by
Guth in the 80's\cite{Guth},
in order to solve some of the shortcomings of the big bang theory, and in
particular, to explain the extraordinary homogeneity of the observable 
universe. However, the universe after inflation in this scenario
becomes very inhomogeneous. Following a detailed investigation
of this problem, A. Guth and E. Weinberg concluded that the old
inflationry model could not be improved\cite{GW}. 

These problems were sorted out by A. Linde in 1983
with the introduction of chaotic inflation\cite{lin}.
In this scenario inflation can occur in theories with
potentials such as $V(\varphi) \sim \varphi^n$. It may
begin in the absence of thermal equilibrium in the early
universe, and it may start even at the Planckian density, in
which case the problem of initial conditions for inflation
can be easily solved\cite{libro}. According to the
simplest versions of chaotic inflationary theory, the 
universe is not a single expanding ball of fire produced in the big
bang, but rather a huge eternally growing fractal. It consists
of many inflating balls that produce new balls, which produce
more new balls, ad infinitum.

BW cosmology has been studied very recently within\cite{....} and without
inflation\cite{u}.
However, inflationary cosmology from the STM theory, still remains without a 
detailed study.
The aim of this paper is the study of inflation from the STM theory of
gravity. More exactly,
I am interested in the study of de Sitter and power-law
expansions for the universe during inflation from
the STM theory developed by P. Wesson\cite{WE}.
This paper is organized as follows. In Sec. II,
I study the STM basic equations. In Sec. III, I develope the classical
and quantum field dynamics during the inflationary stage using the
semiclassical approach developed in \cite{11}. Furthermore,
de Sitter and power-law expansions
for the universe are examined from the STM theory of gravity
and the results are contrasted with experimental data.
Finally, in Sec. IV some final comments are developed.\\

\section{Basic STM equations}

Following the idea suggested by Wesson and co-workers\cite{5}, in this
section I develope 
the induced 4D equation of state from the
5D vacuum field equations, $G_{AB}=0$ ($A,B=0,1,2,3,4$),
which give the 4D Einstein equations $G_{\mu\nu}=
8\pi G \  T_{\mu\nu}$ ($\mu,\nu =0,1,2,3$).
In particular, we consider a 5D spatially isotropic and flat spherically-symmetric
line element
\begin{equation}\label{1}
dS^2 = - e^{\alpha(\psi,t)} dt^2 + e^{\beta(\psi,t)} dr^2 + e^{\gamma(\psi,t)}
d\psi^2,
\end{equation}
where $dr^2 = dx^2+dy^2+dz^2$ and $\psi$ is the fifth coordinate.
We assume that $e^{\alpha}$, $e^{\beta}$ and $e^{\gamma}$
are separable functions of the variables $\psi$ and $t$.
The equations for the relevant Einstein
tensor elements are
\begin{eqnarray}
G^0_{ \  0}& =& -e^{-\alpha} \left[\frac{3 \dot\beta^2}{4} +
\frac{3\dot\beta \dot\gamma}{4}\right] + e^{-\gamma}\left[
\frac{3 \stackrel{\star\star}{\beta}}{2}
+\frac{3\stackrel{\star}{\beta}^2}{2}-\frac{3\stackrel{\star}{\gamma}
\stackrel{\star}{\beta}}{4}\right],\\
G^0_{ \  4} & = & e^{-\alpha}\left[\frac{3\stackrel{\star\cdot}{\beta}}{2}
+\frac{3\dot\beta
\stackrel{\star}{\beta}}{4}
-\frac{3\dot\beta \stackrel{\star}{\alpha}}{4} - \frac{3
\stackrel{\star}{\gamma} \dot\gamma}{4}\right],\\
G^i_{ \  i} & = &- e^{-\alpha} \left[\ddot\beta + \frac{3\dot\beta^2}{4}+
\frac{\ddot\gamma}{2} + \frac{\dot\gamma^2}{4} + \frac{\dot\beta\dot\gamma}{2}-
\frac{\dot\alpha\dot\beta}{2} - \frac{\dot\alpha\dot\gamma}{4}\right] \nonumber \\
& + & e^{-\gamma}\left[\stackrel{\star\star}{\beta}
+\frac{3\stackrel{\star}{\beta}^2}{4}
+ \frac{\stackrel{\star\star}{\alpha}}{2} + \frac{
\stackrel{\star}{\alpha}^2}{4} +
\frac{\stackrel{\star}{\beta}\stackrel{\star}{\alpha}}{2}
-
\frac{\stackrel{\star}{\gamma}\stackrel{\star}{\beta}}{2}
- \frac{\stackrel{\star}{\alpha}\stackrel{\star}{\gamma}}{4}\right],\\
G^4_{ \  4} & = & e^{-\alpha} \left[\frac{3\ddot\beta}{2} +
\frac{3\dot\beta^2}{2}-
\frac{3 \dot\alpha \dot\beta}{4}\right] 
+e^{-\gamma}\left[\frac{3\stackrel{\star}{\beta}^2}{4} +
\frac{3\stackrel{\star}{\beta}\stackrel{\star}{\alpha}}{4} \right],
\end{eqnarray}
where the overstar and the overdot denote respectively ${\partial \over
\partial\psi} $ and ${\partial \over \partial t}$, and
$i=1,2,3$. Following the convention $(-,+,+,+)$ for the
4D metric, we define
$T^0_{ \  0} = -\rho_t$ and $T^1_{ \  1} = {\rm p}$, where
$\rho_t$ is the total energy density and ${\rm p}$ is the pressure.
The 5D-vacuum conditions ($G^A_B =0$) are given by\cite{WE}
\begin{eqnarray}
&& 8\pi G \rho_t = \frac{3}{4} e^{-\alpha} \dot\beta^2, \label{6} \\
&& 8\pi G {\rm p} = e^{-\alpha} \left[\frac{\dot\alpha\dot\beta}{2} -
\ddot\beta - \frac{3\dot\beta^2}{4}\right], \label{7} \\
&& e^{\alpha} \left[\frac{3\stackrel{\star}{\beta}^2}{4} + \frac{3
\stackrel{\star}{\beta}\stackrel{\star}{\alpha}}{4}\right]=
e^{\gamma} \left[\frac{\ddot\beta}{2} + \frac{3\dot\beta^2}{2} - \frac{
\dot\alpha\dot\beta}{4}\right]. \label{8}
\end{eqnarray}
Hence, from eqs. (\ref{6}) and (\ref{7}) and taking
$\dot\alpha = 0$, we obtain
the equation of state for the induced matter 
\begin{equation}\label{9a}
{\rm p} = - \left(\frac{4}{3} \frac{\ddot\beta}{\dot\beta^2} + 1\right)
\rho_t.
\end{equation}
Notice that for $\ddot\beta/\dot\beta^2 \le 0$ and $\left|\ddot\beta/\dot\beta^2
\right| \ll 1$ (or zero), this equation describes an inflationary universe.
The equality $\ddot\beta/\dot\beta^2 =0$ corresponds with a 4D de Sitter
expansion for the universe.

In this paper I explore the possibility to obtain inflation for the
metrics (\ref{1}) with the restrictions
\begin{equation}\label{8a}
\alpha \equiv \alpha(\psi); \quad \beta \equiv \beta(\psi,t);
\quad \gamma \equiv \gamma(t),
\end{equation}
where $e^{\beta}$ is a separable function of $\psi$ and $t$. The conditions
(\ref{8a}) imply that $\dot\alpha = \stackrel{\star}{\gamma}=0$. Furthermore,
in that follows we restrict our study to imflationary models on 5D
metrics where all the coordinates are independent. The choice (\ref{8a})
implies that only the spatial sphere and the the fifth coordinate have
respectively squared sizes $e^{\beta}$ and $e^{\gamma}$ that evolve with
time.

\section{Inflationary universe}

The dynamics of a scalar field minimally coupled to gravity
is described by the Lagrangian density
\begin{equation}
{\cal L}(\varphi,\varphi_{,\mu})=-\sqrt{-g}\left[ \frac{R}{16\pi}+
\frac{1}{2}g^{\mu\nu}\varphi_{,\mu}\varphi_{,\nu} + V(\varphi)\right],
\end{equation}
where $R$ is the 4D scalar curvature and         
$g$ is the determinant of the 4D metric tensor.
If the spacetime has a Friedman-Robertson-Walker (FRW) metric, $
ds^2=-d\tau^2+a^2(\tau) dr^2$, the resulting equations of motion for
the field
operator $\varphi$ and the Hubble parameter, are
\begin{eqnarray}
&&\ddot\varphi-\frac{1}{a^2}\nabla^2\varphi+3H\dot\varphi+V'(\varphi)=0 \ ,\\
&&H^2=\frac{4\pi}{3M_p^2}\left< \dot\varphi^2
+\frac{1}{a^2}(\vec\nabla\varphi)^2 +2V(\varphi) \right> \ ,
\end{eqnarray}
where, in the following the overdot represents the derivative
with respect to $\tau$ and $V'(\varphi)\equiv
\frac{d V}{d\varphi}$. The expectation value is assumed to be a constant
function of the spatial variables for consistency with the FRW metric.  The
discussion of the inflationary stage is usually done through the classical
analysis of the above equations, which in the case of a homogeneous scalar
field and null spatial curvature reduce
to a system of two first order equations, even without any slow-roll
assumption \cite{copeland}. The period in which $\ddot a>0$ and thus the
universe has an accelerated expansion is the inflationary stage, and models
are discarded or not depending on if they provide enough inflation or not.
When the inflation ends the field starts oscillating rapidly and its
potential energy is converted into thermal energy.  This is the general
scheme of the inflationary scenario without considering the quantum effects.
In the usual formulation of this approach the slow-roll regime is assumed.
Instead, we avoid here the use of such an assumption and consider a
consistent semiclassical expansion of the theory. To obtain this we decompose
the inflaton operator in a classical field $\phi_c$ plus a quantum correction
$\phi$, $\varphi=\phi_c+\phi$, such that $<\varphi>=\phi_c$ and
$<\phi> = <\dot\phi>=0$. The field $\phi_c$ is defined as the solution
to the classical equation of motion
\begin{equation}    \label{4}
\ddot \phi_c+3H\dot\phi_c+V'(\phi_c)=0 \ ,
\end{equation}
where we have assumed that the classical field is spatially
homogeneous, in agreement
with the hypothesis of an inflationary regime. The evolution of the quantum
operator $\phi$ is given by
\begin{equation}    \label{5}
\ddot \phi-\frac{1}{a^2}\nabla^2 \phi +3H\dot\phi+\sum_n \frac{1}{n!}
V^{(n+1)}(\phi_c) \phi^n=0 \ .
\end{equation}
At the same time the squared Hubble parameter can be expanded as
\begin{equation}\label{H}
H^2=H^2_c+ \frac{4\pi}{3M^2_p}
\left< \dot \phi^2+\frac{1}{a^2}(\vec\nabla\phi)^2 +
\sum_n \frac{1}{n!} V^{(n+1)}(\phi_c) \phi^n\right> ,
\end{equation}
where
\begin{equation}   \label{7'}
H_c^2=\frac{4\pi}{3M_p^2}\left[\dot \phi_c^2+ 2V(\phi_c)\right],
\end{equation}
is the classical Hubble parameter.  If the quantum fluctuations are small,
hence the inflation is driven by the classical field.
The effect of the classical field $\phi_c$ is to change the
environment in which
the $\phi$ field evolves, and in particular the mass of the $\phi$ particles.
Making a first order expansion in eq. (\ref{H}) for $\phi$, we obtain
$H\equiv H_c=\dot a/a$ in eq.(\ref{5}), which reduces to
\begin{equation}    \label{8'}
\ddot \phi-\frac{1}{a^2}\nabla^2 \phi +3H_c\dot\phi+V"(\phi_c) \phi=0 \ .
\end{equation}
From eqs. (\ref{4}) and (\ref{7'}), we can describe the classical dynamics
of the Hubble parameter and the inflaton field by the relations
\begin{eqnarray}
&&\dot \phi_c =-\frac{M_p^2}{4\pi} H_c'\ , \label{9}\\
&&\dot H_c = H_c'\dot\phi_c =-\frac{M_p^2}{4\pi} (H_c')^2 \ .\label{10}
\end{eqnarray}
In these terms the potential accounting for such a dynamics is given by:
\begin{equation}       \label{11}
V(\phi_c)=\frac{3M_p^2}{8\pi}\left(H_c^2-
\frac{M_p^2}{12\pi}(H_c')^2\right) \label{p2} \ .
\end{equation}
Equations (\ref{9}) and (\ref{10}) define the classical evolution of
spacetime, determining the relation between the classical potential and the
inflationary regimes. On the other hand eq. (\ref{8}) defines the quantum
dynamics of the field $\phi$, whose classical character was studied
in\cite{11,12}. In this framework, the total energy density and the
pressure (neglecting the terms $\left<\dot\phi^2\right>/2$,
because they are very small on cosmological scales), are given by
\begin{eqnarray}
&& \left< \rho_t\right> = \frac{\dot\phi^2_c}{2} + V(\phi_c), \\
&&\left<{\rm p} \right> = \frac{\dot\phi^2_c}{2} - V(\phi_c),
\end{eqnarray}
such that the equation of state for supercooled inflation is
\begin{equation}\label{b}
\left< {\rm p} \right> = -\left(\frac{2 \dot H_c}{3 H^2_c}+1 \right)
\left<\rho_t\right>.
\end{equation}
From eqs. (\ref{9a}) and (\ref{b}) one obtains the Hubble parameter as a
function of $\beta$ for inflationary models with
$\stackrel{\star}{\gamma} = \dot\psi =0$
\begin{equation}\label{h}
H_c = \dot\beta/2.
\end{equation}
Furthermore, from eqs. (\ref{h}) and (\ref{11}) we can write the scalar
potential for inflationary models that follows from the 5D metric
(\ref{1})
\begin{equation}\label{v}
V[\phi_c(t)] = \frac{3 M^2_p}{2\pi} \left[ \dot\beta^2 +
\frac{2}{3}\ddot\beta\right].
\end{equation}
The equation (\ref{h}) shows that models with $H_c = {\rm const.}$ (like
a de Sitter one) are described with potentials
\begin{equation}\label{w}
V[\phi_c(t)] = \frac{3 M^2_p}{2\pi} \dot\beta^2.
\end{equation}

To map the eq. (\ref{8}) in a wave equation, we can make the transformation
$\chi =e^{\frac{3}{2}\int d\tau H}\phi= a^{\frac{3}{2}}\phi$
\begin{equation}
\ddot \chi -\frac{1}{a^2}\nabla^2\chi -\frac{k_0^2}{a^2}\chi=0 \ ,
\end{equation}
where $k_0^2 = a^2 \left(\frac{9}{4}H^2_c+\frac{3}{2}\dot H_c-V"_c\right)$. We
now have a scalar field with a time-dependent mass. It can be described
in terms of a set of modes
\begin{equation}   \label{13}
\chi=\frac{1}{(2\pi)^{\frac{3}{2}}}\int d^3k \left[a_k e^{i\vec k.\vec r} \xi_k(t)
+ h.c.\right] \ ,
\end{equation}
where we have made explicit use of the spatial homogeneity of the FRW metric.
The operators $a_k$ and $a_k^\dagger$ satisfy the canonical commutators:
\begin{eqnarray}
&&[a_k,a_{k'}^\dagger]= \delta(\vec k-\vec k') \ , \label{14}\\
&&[a_k,a_{k'}]=[a_k^\dagger, a_{k'}^\dagger]=0 \ , \label{15}
\end{eqnarray}
and the equation of motion for the modes $\xi_k(t)$ are given by
\begin{equation}    \label{16}
\ddot\xi_k + \omega_k^2 \xi_k =0 \ ,
\end{equation}
with $\omega_k^2=\frac{1}{a^2}\left(k^2 - k_0^2\right)$. Here, 
$k_0(\tau)$ separates the large scale (unstable)
sector ($k^2<k_0^2$) and the short scale (stable) one
($k^2>k_0^2$).  When $k_0$
surpasses $k$, the temporal oscillation of the mode ceases. During the
expansion of the universe, more and more new fluctuations suffer this
transformation.  These quantum fluctuations with wave number below $k_0$ are
generally interpreted as inhomogeneities superimposed to the classical field
$\phi_c$. They are
responsible for the density inhomogeneities generated during the
inflation.

In order that $\chi$ and $\dot \chi$ satisfy a canonical commutation
relation, $[\chi(\vec r,t), \dot \chi(\vec{r'},t)] = i\delta(\vec r-\vec{r'})$,
the normalization of the modes $\xi_k(t)$ must be chosen
\begin{equation} \label{17}
\xi_k\dot\xi_k^*-\dot\xi_k\xi_k^*=i,
\end{equation}
where the asterisk denotes the complex conjugate.
In the following subsections I will study two particular cases of
inflation from the STM theory of gravity;
de Sitter and power-law expansions.

\subsection{de Sitter expansion}

The special case $e^{\alpha}=\psi^2$ in eq. (\ref{v}) [see eqs.
(\ref{h}) and (\ref{w})], with $\ddot\beta/\dot\beta^2=0$,
$\stackrel{\star}{\gamma}=\dot\gamma =0$ in eq. (\ref{1}),
gives a de Sitter expansion
for which $<\rho_t+{\rm p}> = 0$, so that  $\phi_c=\phi_0$, $V(\phi_c)=V_0$
and $H_c(\tau) = H_0$ are constant\cite{11}. It implies that
\begin{equation}
H^2_0= \frac{ 8 \pi}{3 M^2_p} V_0 \ .
\end{equation}
This case
corresponds exactly to a scalar field $\phi$ with a mass
$m^2=\left(\frac{d^2V}{d\phi^2}\right)_{\phi=\phi_0}$ in a de Sitter
background with a constant Hubble parameter $H_0$ and a scale factor
$a(t) \sim e^{H_0 \tau}$. For this model 
the cosmological constant $\Lambda$ gives the vacuum energy density
\begin{equation}\label{a1}
<\rho_t> = \frac{\Lambda}{8\pi G},
\end{equation}
such that $\Lambda$ is related with the fifth coordinate by means of
\begin{equation}
\Lambda = 3/\psi^2.
\end{equation}
The 5D-metric for a de Sitter expansion of the universe is\cite{1}
\begin{equation} \label{a2}
dS^2 = -d\tau^2 + \psi^2 e^{2\sqrt{\Lambda/(3\psi^2)}  \tau} dr^2 + d\psi^2.
\end{equation}
Hence, the 4D de Sitter inflationary universe can be seen as a 5D metric
with the fifth constant coordinate and unitary size. The equation (\ref{a1})
shows that the vacuum on the metric (\ref{a2}) induces the cosmological
constant $\Lambda$, which depends on the
value of the fifth coordinate and generates
the expansion of the universe. 

On the other hand, the wave number $k_0$ is
\begin{equation}
k_0 = H_0 a(\tau)\sqrt{\frac{9}{4} -\frac{m^2}{H_0^2}} \ .
\end{equation}
The equation of motion for the modes in a de Sitter expansion
can be written explicitely
\begin{equation}
\ddot \xi_k(\tau) + \left[k^2 e^{-2H_0 \tau}
- \frac{9}{4} H^2_0 + m^2 \right] \  \xi_k(\tau)=0 \ ,
\end{equation}
and its general solution is
\begin{equation}
\xi_k(\tau) = A_1
{\cal H}^{(1)}_{\nu}\left[\frac{k}{ H_0 a}\right] +
A_2 {\cal H}^{(2)}_{\nu}\left[\frac{k}{ H_0 a}\right] \ ,
\end{equation}
where ${\cal H}^{(1)}_{\nu}$ and ${\cal H}^{(2)}_{\nu}$ are the 
Hankel functions and $\nu=\sqrt{{9 \over 4}
-\frac{m^2}{H^2_0}}<\frac{3}{2}$.
The long time behavior of the mean square of $\phi$ reproduces the exact
value as calculated by standard quantum field methods\cite{habib}. For a
massive inflaton (i.e., with $ m/H_0 \ll 1$), once
we taken into account the normalization condition for the modes (\ref{17}),
we have
\begin{equation}
\left.\left<\phi^2\right>\right|_{SH}\simeq
\frac{a^{-(3-2\nu)}(\tau)}{2^{3-2\nu} \pi^3} \Gamma^2(\nu) H^{2\nu-1}_0
{\Large\int}^{k_0}_{0} \frac{dk}{k} k^{3-2\nu},
\end{equation}
where $\Gamma(\nu)$ is the gamma function with $\nu$-argument.
Hence, the scale invariant spectrum $n_s = 1$ (i.e., $\nu = 1/2$) implies
that $m^2 = 2 H^2_0 = 2/(3\psi^2)$, and
\begin{equation}\label{f}
\left.\left<\phi^2\right>\right|_{SH}\simeq
\frac{\Gamma^2(1/2)}{24 \pi^3} \psi^{-2},
\end{equation}
where $\Gamma$ is the gamma function.
Hence, the fifth coordinate also should be related to the inflaton
fluctuations for a scale invariant power spectrum.

\subsection{Power-law inflation}

To study a power-law expansion of the universe, we can make
$e^{\alpha} = \psi^s$, $e^{\beta} = a^2(t) \psi^r$ and
$e^{\gamma} = A \  t^n$, in eq. (\ref{1}). Hence, the 5D-line element
can be written as
\begin{equation}
dS^2  = - \psi^s dt^2 + a^2(t) \psi^r dr^2 + A t^n d\psi^2,
\end{equation}
where 
the scale factor is $a \sim t^p$. From the 5D-vacuum conditions
(6-8),
we find
\begin{equation}
s=n=2, \quad r=\frac{2p}{p-1}, \quad A=(p-1)^{-2},
\end{equation}
such that we obtain the Ponce de Leon metric\cite{1}
\begin{equation}\label{11''}
dS^2 = -\psi^2 dt^2 + a^2(t) \psi^{\frac{2p}{p-1}} dr^2 +
(p-1)^{-2} t^2 d\psi^2.
\end{equation}
This metric is an exact solution of the 5D field equations in apparent
vacuum, which in terms of the Ricci tensor are $R_{AB}=0$. The induced
metric $h_{\alpha\beta}$ on these hypersurfaces is given by
\begin{equation}
ds^2 = h_{\alpha\beta} dx^{\alpha} dx^{\beta} = - d\tau^2+ a^2(\tau) dr^2,
\end{equation}
where $\tau = \psi t$ is the cosmic time which
corresponds to the spatially flat FRW metric.
With this representation, the scale factor is
\begin{equation}
a(\tau) = \psi^{\frac{p(2-p)}{p-1}} \  \tau^p.
\end{equation}
I am interested in the study of power-law inflationary dynamics from
the metric (\ref{11''}).
In this case $p>1$, and the equation of state if matter is interpreted
as a perfect fluid, is
\begin{equation}
\left<{\rm p}\right> = -\left(\frac{3p-2}{3p}\right) \left<\rho_t\right>.
\end{equation}
From eqs. (\ref{10}) and (\ref{v}), one obtains
\begin{equation}
\left(H'_c\right)^2 = - \frac{2\pi}{M^2_p} \ddot\beta.
\end{equation}
Replacing in eq. (\ref{11}) the scalar potential can be obtained
as a function of $\beta(\psi,t)$
Furthermore the constant $p$ is related with the properties of
matter
\begin{equation}\label{alpha}
p = \frac{2}{3} \left(\frac{<\rho_t>}{<\rho_t+{\rm p}>}\right),
\end{equation}
and the temporal evolution for $\phi_c(\tau)$ can be obtained
from eq. (\ref{9})
\begin{equation}\label{beta}
\phi_c(\tau) = - \frac{M_p}{2} \sqrt{\frac{p}{\pi}} {\rm ln}\left[
\frac{H_0}{p} \tau\right].
\end{equation}
Hence, due to $\tau = \psi t$, we obtain the value of the constant $\psi$
\begin{equation}          \label{gamma}
\psi = \frac{p}{H_0 t_0} e^{-\frac{2}{M_p} \sqrt{\frac{\pi}{p}} \phi_0},
\end{equation}
where ($H_0,t_0, \phi_0$) are respectively the values of
($H_c,t,\phi_c$) when inflation starts.
This implies that $\psi$ should be strongly dependent of initial conditions.
Note that $p$ is related to the properties of matter such as
the energy density $\rho_t$ and pressure ${\rm p}$. In the
induced-matter interpretation of KK theory, these are the result of the
geometry in five dimensions, insofar as they are functions of the extra
part of the metric and partial derivatives of the metric coefficients with
respect to the extra coordinate. The field equations of the theory in terms
of the Ricci tensor are $R_{AB}=0$. These appear to describe empty
five-dimensionnal space. But they in fact contain the Einstein equations
$G_{\alpha\beta} = 8\pi G T_{\alpha\beta}$.

Finally, taking into account the results above, the Ponce de Leon
metric (\ref{11''}) can be written as
\begin{equation}\label{nue}
dS^2 = - d\tau^2 + \psi^{\frac{2p}{p-1}} \left(\frac{p}{H_0 \psi}\right)^{2p}
e^{-4\frac{\sqrt{\pi p }}{M_p} \phi_c(\tau)} dr^2 +
\left(\frac{p}{p-1}\right)^2 \left(\frac{1}{H_0\psi}\right)^2
e^{-4\sqrt{\frac{\pi}{p}} M^{-1}_p \phi_c(\tau)} d\psi^2,
\end{equation}
where ($p,\phi_c,\psi$) are given respectively by eqs. (\ref{alpha}),
(\ref{beta}) and (\ref{gamma}).
Notice the $\phi_c(\tau)$-dependence
of the metric (\ref{nue}), which make the difference between power-law
inflationary scenarios and other cosmological models with $p <1$,
studied in \cite{1}.
From eq. (\ref{nue}) can be view that the size of the fifth coordinate
grows with $\tau $, but the 5D vacuum here requiered implies that $\psi =
{\rm const.}$. Hence, for a power-law inflationary expansion, the 4D
universe that expands from a metastable 4D vacuum with a scalar potential
given by eq. (\ref{v}), is realy a 5D universe in a true vacuum with a metric
given by eq. (\ref{nue}).
The differential equation (\ref{16}), has the solution
\begin{equation}
\xi_k(\tau) = A_1 \sqrt{\frac{\tau}{\tau_0}}
{\cal H}^{(1)}_{\nu}\left[\frac{k \tau^{1-p} H_0}{(p-1)\tau^{-p}_0}\right]
+ A_2 \sqrt{\frac{\tau}{\tau_0}} {\cal H}^{(2)}_{\nu}\left[\frac{k
\tau^{1-p} H_0}{(p-1)\tau^{-p}_0}\right],
\end{equation}
where ${\cal H}^{(1)}_{\nu}$ and ${\cal H}^{(2)}_{\nu}$ are the
Hankel functions and
$\nu = \sqrt{{9 \over 4}p^2 - {15\over 2} p + {9\over 4}}/(p-1)$.
From the normalization condition (\ref{17}), we obtain the adiabatic
vacuum
\begin{equation}
\xi_k(\tau) = \sqrt{\frac{\pi}{2}} \sqrt{\frac{\tau}{(p-1)\tau_0}}
{\cal H}^{(2)}_{\nu} \left[\frac{k \tau^{1-p} H_0}{(p-1) \tau^{-p}_0}\right].
\end{equation}
The Hankel functions take the small-argument
limit $\left.{\cal H}^{(2,1)}_{\nu}[x]\right|_{x\ll 1} \simeq
{(x/2)^{\nu} \over \Gamma(1+\nu)} \pm {i\over \pi}
\Gamma(\nu) \left(x/2\right)^{-\nu}$.
Hence, the super Hubble
squared fluctuations $\left.\left<\phi^2\right>\right|_{SH}$ are
\begin{equation}\label{ii}
\left.\left<\phi^2\right>\right|_{SH}
\simeq \frac{4^{\nu-1} \Gamma^2(\nu) (p-1)^{2\nu-1}}{\pi H^{3+2\nu}_0}
\frac{\tau^{-p(3-2\nu)-2\nu+1}}{\tau^{-p(3-2\nu)+1}_0}
{\Large \int}^{k_0}_{0} \frac{dk}{k} k^{3-2\nu},
\end{equation}
where $k_0(\tau) = \psi^{p(2-p) \over p-1}
\left[{9\over 4}p^2-{15\over 2}p+1\right]^{1/2}
\tau^{1-p}$ and $\Gamma(\nu)$ is the gamma function with $\nu$-argument.
The integral controlling the presence of infrared divergences is
${\Large\int}^{k_0(t)}_0 dk \  k^{2(1-\nu)}$, which has
a power spectrum ${\cal P}_{<(\delta\phi)^2>} \sim k^{3-2\nu}$.
We can evaluate the expression (\ref{ii}), and we obtain
\begin{equation}
\left.\left<\phi^2\right>\right|_{SH} \simeq
\frac{4^{\nu-1} \Gamma^{2}(\nu) \left[\frac{9}{4}p^2-\frac{15}{2}p+1\right]^{
\frac{(3-2\nu)}{2}} \left( p-1\right)^{2\nu-1} \psi^{\frac{p(2-p)(3-2\nu)}{p-1}}}{
\pi H^{3+2\nu}_0 \tau^{1-p(3-2\nu)}_0 \left(3-2\nu\right)} \  \tau^{1-2\nu}.
\end{equation}
The spectral index $n_s=3/2-\nu \simeq 1$ agrees
with the experimental data\cite{prl} for $\nu \simeq 1/2$ (i.e., for
$p \simeq 3.08$). For this value of $\nu$, we obtain a time independent
$\left.\left<\phi^2\right>\right|_{SH}$, so that the SH metric fluctuations
in (\ref{nue}) remains constant for a scale invariant power spectrum.

\section{Final Comments}

In this work we studied  the evolution of the early universe
from the
STM theory of gravity.
In this framework
inflation should be a consequence of the expansion generated
by a cosmological ``constant'' (variable in power-law inflation and
constant in a de Sitter expansion) induced by a generalized 5D spatially
flat, isotropic an homogeneous background metric where the fifth dimension
becomes constant as a consequence of the 5D vacuum state. The condition
to obtain a inflationary expansion is
\begin{displaymath}
\left<{\rm p}\right> = - \left(\frac{4}{3} \frac{\ddot\beta}{\dot\beta^2} +
1\right) \left<\rho_t\right> \simeq - \left<\rho_t\right>,
\end{displaymath}
or respectively
\begin{displaymath}
\frac{\ddot\beta}{\dot\beta^2} \ll 1, \qquad
\frac{\ddot\beta}{\dot\beta^2} =0,
\end{displaymath}
for power-law and de Sitter expansions where $\dot\beta = 2 H_c$.

In the models here studied all of the components of the Riemann-Christoffel
tensor for the 5D metric are zero. Despite this, the model's 4D
part is not flat, since the 4D Ricci scalar is non zero. Hence, we see
that while the inflationary universe may be curved in 4D, it is flat in 5D.
Thus, the 5D STM theory of gravity describes an inflationary universe
consisting of localized singular sources embedded in a globally 5D flat
cosmology.

We have restricted our study
to 4D spatially flat FRW metrics, which are relevant to inflationary
cosmology. We studied the power-law and de Sitter models of inflation.
We find that
$a)$ for a de Sitter
expansion the situation appears to be different. In this case the additional
dimension is related to the cosmological constant $\Lambda = 3/\psi^2$ and
the Hubble parameter is $H_0 = 1/\sqrt{3\psi^2}$. Hence, we inquire that
$\psi$ has $G^{1/2}$ unities. From the experimental data 
obtained from BOOMERANG-98, MAXIMA-1 and COBE DMR\cite{prl}, we obtain
that $m^2 \psi^2=2/3$. This implies that, in a de Sitter expansion
with a 5D vacuum state, $\psi$ gives nearly
the inverse of the mass of the inflaton field. Hence, for $m^2 \simeq
(10^{-8} - 10^{-12}) \  M^2_p$, the resulting fifth coordinate will
be of the order of $\psi^2 \simeq (10^8 - 10^{12}) \  M^{-2}_p$.
Note that the inflaton field fluctuations are also related to the fifth
coordinate [see eq. (\ref{f})]. Hence, in the induce-matter interpretation
of KK theory, matter field fluctuations also appear as a consequence
of the five dimensional geometry.
$b)$ For the power-law expansion, the fifth coordinate appears
to be a dimensionless constant, which, in principle, could be considered
as 1 (notice that we are considering $c=\hbar=1$). The interesting here is
that the $\psi$-value depends on the initial conditions during power-law
inflation [see eq. (\ref{gamma})]. Furthermore, the main result of this
paper is the metric (\ref{nue}) which describes the 5D line element
in power-law inflation. Notice the dependence with
the spatially isotropic component of the
inflaton field, $\phi_c$. Furthermore, it is very important the fact that
the squared size of the fifth coordinate $e^{\gamma}$ grows with
time in a power-law inflationary universe.\\

\vskip .05cm
\centerline{{\bf ACKNOWLEDGMENTS}}
\vskip .05cm
\noindent
I would like to acknowledge CONICET (Argentina) and Universidad
Nacional de Mar del Plata for financial
support in the form of a research grant.


\begin{thebibliography}{99}
\bibitem{1} J. Ponce de Leon, Gen.Rel. Grav. {\bf 20}, 539 (1988).
\bibitem{2} P. S. Wesson, Phys. Lett. {\bf B276}, 299 (1992).
\bibitem{2'} P. S. Wesson, Astrophys. J. {\bf 394},19 (1992);
Astrophys. J. {\bf 436}, 547 (1994);
Astrophys. J. {\bf 440}, 1 (1995).
\bibitem{3} D. J. Gross and M. J. Perry, Nucl. Phys. {\bf B226}, 29
(1983).
\bibitem{4} A. Davidson and D. Owen, Phys. Lett. {\bf B155}, 247 (1985).
\bibitem{.} N. Arkani-Hamed, S. Dimopoulos and G. Dvali, Phys.
Lett.{\bf B429}, 263 (1998); I. Antoniadis, N. A. Arkani-Hamed, S. Dimopoylos
and G. Dvali, Phys. Lett. {\bf B436}, 257 (1998); L. Randall and R. Sundrum,
Phys. Rev. Lett. {\bf 83}, 3370 (1999); L. Randall and R. Sundrum, Phys.
Rev. Lett. {\bf 83}, 4690 (1999).
\bibitem{..} R. Kallosh and A. Linde, JHEP {\bf 0002}, 005 (2000);
K. Behrndt and M. Cvetic, Phys. Rev. {\bf D61}, 101901 (2000); G. W.
Gibbons and N. D. Lambert, Phys. Lett. {\bf B488}, 90 (2000);
J. Maldacena and C. Nunez, Int. J. Mod. Phys. {\bf A16}, 822 (2001).
\bibitem{...} S. S. Seahra and P. S. Wesson,
Class. and Quant. Grav. {\bf 19}, 1139 (2002).
\bibitem{....} S. Nojiri and S. D. Odintsov, JHEP {\bf 0112}, 033 (2001).
\bibitem{u} R. Mansouri and F. Nasseri, Phys. Rev. {\bf D60},
123512.
\bibitem{5} P. Wesson, Gen. Rel. Grav. {\bf 22}, 707 (1990);
Astrophy. J. {\bf 394}, 19 (1992); P. Wesson, H. Liu and P. Lim,
Phys. Lett. {\bf B298}, 69 (1993).
\bibitem{1'} A. A. Starobinsky,
JETP Lett. {\bf 30}, 682 (1979); Phys. Lett. {\bf B91},
99 (1980).
\bibitem{Guth} A. Guth, Phys. Rev. {\bf D23}, 347 (1981).
\bibitem{GW} A. Guth, E. J. Weinberg, Nucl. Phys. {\bf B212}, 321 (1983).
\bibitem{lin} A. D. Linde, Phys. Lett. {\bf 129}, 177 (1983).
\bibitem{libro} For a review of inflation the reader can see A. D.
Linde, {\em Particle Physics and Inflationary Cosmology}, Harwood,
Chur, Switzerland, 1990; A. D. Linde, Phys. Rep. {\bf 333},
575 (2000).
\bibitem{WE} For a review of STM theory the reader can see
J. M. Overduin and P. S. Wesson, Phys. Rep. {\bf 283}, 303 (1997).
\bibitem{11} M. Bellini, H. Casini, R. Montemayor and P. Sisterna,
Phys. Rev.{\bf D54}, 7172 (1996).
\bibitem{copeland} E.J.Copeland, E.W.Colb, A.R.Liddle and J.E.Lidsay, Phys.
Rev. {\bf D48}, 2529 (1993).
\bibitem{12} H. Casini, R. Montemayor and P. Sisterna, Phys. Rev. {\bf D59},
063512 (1999).
\bibitem{prl} A. H. Jaffe et. al, Phys. Rev. Lett. {\bf 86}, 3475
(2001).
\bibitem{habib} S.Habib, Phys. Rev. {\bf D46}, 2408 (1992).
\end{thebibliography}
\end{document}